# An ultraviolet photodetector based on conductive hydrogenated TiO$_2$ film prepared by radio frequency atmospheric pressure plasma

*Yu Zhang, Haozhe Wang, Jie Cui, Tao He, Gaote Qiu, Yu Xu\*, and Jing Zhang\**

Y. Zhang, H. Wang, J. Cui, T. He, G. Qiu
College of science, Donghua University, Shanghai 201620, China.
Y. Xu, J. Zhang
College of science, Donghua University, Shanghai 201620, China; Textile Key Laboratory for Advanced Plasma Technology and Application, China National Textile & Apparel Council, Shanghai 201620, China; Magnetic Confinement Fusion Research Center, Ministry of Education, Shanghai 201620, China.
Email: yuxu@dhu.edu.cn (Y.X.) and jingzh@dhu.edu.cn (J.Z.)

Keywords: conductive TiO$_2$ film, Ti$^{3+}$/Ti$^{2+}$ self-doped, UV photodetector, atmospheric pressure plasma, photo-response

Abstract: The growing demand for real-time ultraviolet (UV) monitoring calls for a simple, rapid, and low-cost strategy to prepare UV photodetectors. We prepare a wearable real-time UV photodetector based on hydrogenated titanium dioxide film synthesized by radio frequency atmospheric pressure plasma. The conductivity of our hydrogenated titanium dioxide is improved to 10.2 S cm$^{-1}$, 9 orders of magnitude higher than that of pristine titanium dioxide after 10 min plasma treatment. Plasma hydrogenation disrupts the surface crystal structure, introducing oxygen vacancies (OVs) that create self-doped titanium(III) and titanium(II) species. First-principles calculations indicate the OVs raise the Fermi level of TiO$_2$ and distort the lattice nearby. Our optimized film has a distinctive periodic switching characteristic under intermittent illumination and good responsivity from 280–400 nm, peaking at 632.35 mA W$^{-1}$ at 365 nm. The fabricated wearable sensor based on the optimized film effectively performs the monitoring of the daily variation of ambient UV intensity in three typical weather types and transferring its data to a smartphone via Wi-Fi.

## 1. Introduction

Ultraviolet (UV) radiation has an outsize impact on the evolution and survival of life on Earth, even though it only accounts for a small portion–about 10%–of solar radiation.[1] The UV spectrum is subdivided into 4 regions: UV-A (320–400 nm), UV-B (280–320 nm), UV-C (200–280 nm), and vacuum UV (10–200 nm).[1b, 2] The ozone layer filters UV-C and vacuum UV radiation, but UV-A and UV-B still penetrate to the ground,[3] so everyone is irradiated by UV in daily life. Moderate UV radiation is believed to be beneficial to human health and is essential for producing vitamin D. Excessive UV irradiation, however, induces both acute and chronic damage to the skin, eyes, and immune system. It also ravages the productivity of crops and the lifespan of buildings. Therefore, the detection of and protection from UV is crucial in myriad applications, including outdoor sports [1c, 4], chemistry [5], biology [4, 6], medicine [1c], astronomical exploration [1c, 5a], biomedical imaging [1c, 4-5], security [3a, 4-5], night vision [1c, 4-5], and optical communication [1c, 3a, 4-5].

In recent decades, UV photodetectors (PDs) have shown promise for the practical detection of ultraviolet light.[7] Wide-bandgap semiconductors such as GaN, ZnO, SiC, and $Ga_2O_3$[8] are typically selected as the photosensitive material for UV PDs. $TiO_2$, another wide-bandgap semiconductor, has also been applied in UV PDs. It has many advantages, such as rich mineral storage, chemical stability, low cost, and non-toxicity.[1d, 3a] Pristine $TiO_2$ suffers from low quantum efficiency, but this can be improved by nanocrystalline synthesis, surface modification, and heterojunction construction.[1d] Various novel UV PDs based on $TiO_2$ nanomaterials have been developed [3a, 9], but they generally involve an expensive or complex fabrication processes, making them not easy for large-scale applications. On the other hand, $TiO_2$ nanomaterials would be combined with some conductive materials, such as graphene[9a], zinc oxide[10] and CuZnS[3a], rather than making $TiO_2$ conductive in these strategies. Thus, a simple, rapid, efficient, and low-cost strategy for PD preparation is needed badly.

Hydrogenation is easy as to implement and thus a common strategy for heterojunction construction and conductivity increasing. Cronemeyer[11] found that $TiO_2$ (rutile) reduced with hydrogen at 700°C had the increased conductivity. Chen et al.[12] hydrogenated anatase $TiO_2$ at high pressure; the resulting hydrogenated titanium dioxide (H–$TiO_2$) exhibited enhanced visible light absorption, improved photocatalytic degradation of MB and phenol, and excellent hydrogen ($H_2$) production. Lu et al.[13] annealed various $TiO_2$ at 450°C for 1 h in a 5% $H_2$ and 95% Ar atmosphere (100 sccm flow rate). The hydrogenated $TiO_2$ (H–$TiO_2$) exhibited an increase of conductivity nearly 30 times. Hydrogen plasma as a green technology was also applied in the preparing of H–$TiO_2$. It does not require catalysts or other reagents and the product does not require any separation or purification. Wang et al.[14] prepared H–$TiO_2$ by thermal plasma at 200 W for 4–8 h at 500 °C. Teng et al.[15] performed the hydrogenation of $TiO_2$ in a hot-filament plasma enhanced chemical vapor deposition (PECVD) apparatus for 3 h at 350–500°C. All the H–TiO$_2$ had black color and enhanced absorption in the visible light region. Yan et al.[16] also prepared H–TiO$_2$ by a H$_2$ plasma (inductively coupled plasma, 26.5–28.3 mTorr, 3000 W) at 390°C in 50 sccm H$_2$ flow rate for 3 h. However, they usually require a good sealing pressure chamber and a long reaction time. Radio-frequency (RF) atmospheric pressure (AP) plasma[17] overcomes these problems requiring neither compressure chambers nor pumps. Our group[18] hydrogenated TiO$_2$ films via RF AP plasma, and the H–TiO$_2$ films presented enhanced absorption in the visible light region and good MB degradation ability. Hydrogen plasma can effectively introduce OVs into the TiO$_2$ lattice.[14, 19] Oxygen vacancies increase the concentration of Ti$^{3+}$ and Ti$^{2+}$ species,[18] which play critical roles in determining the surface morphology, electronic properties and optical characteristics of the material.[20] Theoretical and experimental results show that OVs improve the conductivity of TiO$_2$ by raising the Fermi level.[20a, 21] OV-doped TiO$_2$ (H–TiO$_2$) possesses this modulated electron band structure, which yields improved conductivity, charge transport, and photo-response properties. Thus, RF AP hydrogen plasma is a simple, rapid, efficient and low-cost strategy for preparation of H–TiO$_2$ films based UV PD.

In this work, we fabricate a UV PD based on a conductive H–TiO$_2$ film, and use it to monitor ambient UV intensity. This preparation method is fast, environmentally friendly, and energy saving. The H–TiO$_2$ film was deposited directly on a quartz substrate with the He/O$_2$/TiCl$_4$ deposition system, then hydrogenated by RF AP hydrogen plasma. The conductivity and mobility of this film are investigated. It presents significantly improved conductivity (10.2 S cm$^{-1}$), 9 orders of magnitude higher than that of pristine TiO$_2$. The UV-Vis absorbance spectra, morphologies, crystalline structures, chemical components and photoelectric characteristics of H–TiO$_2$ films were also investigated and discussed. We also perform density functional theory (DFT) calculations, which support the hypothesis that OVs raise the Fermi level and increase the conductivity. Finally, the current under UV illumination increases remarkably and the optimized film shows a distinctive periodic switching characteristic under intermittent illumination and good responsivity. The PD monitors effectively the daily variations in ambient UV intensity in three typical weather conditions. This work paves the way through facile, rapid and low-cost plasma treatment for UV monitoring of wide bandgap semiconductor materials.

## 2. Results and discussion
### 2.1. Electric conductivity analysis

We prepared H–TiO$_2$ films on quartz substrate by RF AP plasma at 40, 60, 80, 100, 200 W. H–TiO$_2$-200W exhibits a remarkably high conductivity, $\sigma = 10.2$ S cm$^{-1}$, while H–TiO$_2$-40W has a much lower conductivity, $\sigma = 1.14\times10^{-2}$ S cm$^{-1}$. The conductivity of pristine TiO$_2$ is below the limit of detection ($10^{-5}$ S cm$^{-1}$).[22] Thus, the resistivity of H–TiO$_2$-200W is at least 9 orders of magnitude less than that of pristine TiO$_2$. We attribute this enhancement in conductivity (equivalently, reduction in resistance) in hydrogenated TiO$_2$ to the presence of defects, which effectively act as electron donors.[23]

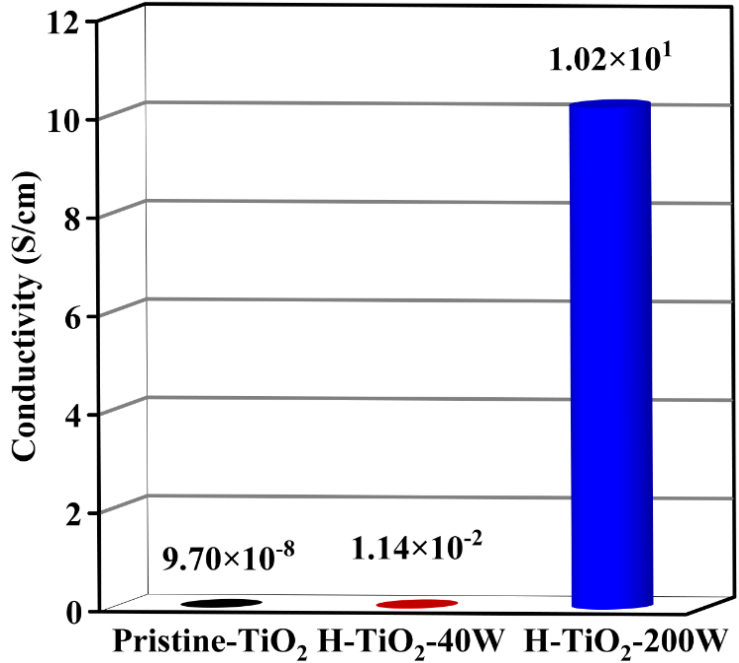

**Figure 1.** Comparing the conductivities of pristine and H–TiO$_2$ films. Pristine TiO$_2$ data from Su et al.[22]

The electric properties of TiO$_2$ films—including the Hall coefficient $R_H$, the carrier concentration $n$, and the mobility $\mu$—are sensitive to the concentration of oxygen vacancies.[24] We calculate them as[25]

$$V_H = R_H \frac{IB}{d}, \tag{1}$$

$$n = \frac{f}{R_H e}, \tag{2}$$

$$\sigma = ne\mu, \tag{3}$$

where $V_H$ is the Hall voltage [V], $I$ is the current [A], $B$ is the magnetic field [T], $d$ is the film thickness [cm], $e = 1.6 \times 10^{-19}$ C is the elementary charge, and $f$ is the Hall scattering factor (typically $1 \leq f \leq 2$, here we assume $f = 1$). The corresponding results are listed in Table 1. The pristine TiO$_2$ film has a negative $R_H$ of $-4.16\times10^4$ cm$^3$ C$^{-1}$, indicating a $n$-type semiconducting character with electron density $n = 1.50\times10^{14}$ cm$^{-3}$. H–TiO$_2$–40W has a smaller negative resistance $R_H = -6.19\times10^3$ cm$^3$ C$^{-1}$, with the electron density $n = 1.01\times10^{15}$

cm$^{-3}$. In comparison, H–TiO$_2$–200W has the smallest negative Hall coefficient, $R_H$ = – 2.16×10$^{-2}$ cm$^3$ C$^{-1}$, and the highest carrier concentration, $n$ = 2.89×10$^{20}$ cm$^{-3}$. The enhanced electron density $n$ leads to more filled electron traps, suppressing the rate of trap-assisted recombination in the film.[25] The mobility does not have a straightforward trend. In pristine TiO$_2$, $\mu$ = 4.04×10$^{-3}$ cm$^2$ V$^{-1}$ s$^{-1}$; TiO$_2$ hydrogenated at 40 W has a larger mobility, $\mu$ = 7.05×10$^{1}$ cm$^2$ V$^{-1}$ s$^{-1}$. However, H–TiO$_2$-200W has $\mu$ = 2.21×10$^{-1}$ cm$^2$ V$^{-1}$ s$^{-1}$, a smaller value. We hypothesize that some hydrogenation of comporrer increases $\mu$ because of improvement in $R_H$, while hydrogenation at higher power decreases $\mu$ by increasing the concentration of defects $n$, which act as charge carrier scattering centers.[26] It shows a "^" shaped trend across the plasma power.

**Table 1.** Summary of Hall effect measurements in pristine and hydrogenated TiO$_2$ films. All act as $n$-type semiconductors.

| Sample | σ [S cm$^{-1}$] | $R_H$ [cm$^3$ C$^{-1}$] | $n$ [cm$^{-3}$] | $\mu$ [cm$^2$ V$^{-1}$ s$^{-1}$] |
| --- | --- | --- | --- | --- |
| Pristine TiO$_2$ | 9.70×10$^{-8}$ [22] | -4.16×10$^{4}$ | 1.50×10$^{14}$ | 4.04×10$^{-3}$ |
| H–TiO$_2$-40W | 1.14×10$^{-2}$ | -6.19×10$^{3}$ | 1.01×10$^{15}$ | 7.05×10$^{1}$ |
| H–TiO$_2$-200W | 1.02×10$^{1}$ | –2.16×10$^{-2}$ | 2.89×10$^{20}$ | 2.21×10$^{-1}$ |

## 2.2. UV-Vis absorption spectra

In Figure 2a, we display photographs of pristine and hydrogenated TiO$_2$ films. H–TiO$_2$-40W maintains its original white color, while H–TiO$_2$-60W, H–TiO$_2$-80W and H–TiO$_2$-100W exhibit a slight grayish hue; H–TiO$_2$-200W is black in color. Figure 2b illustrates their UV-Vis absorption spectra. Absorption in the visible region (400–800 nm) gradually increases with increased plasma power; the absorption edge is also redshifted, which is responsible for the progressive color change from white to black. The bandgap E$_g$ of pristine and H–TiO$_2$ films can be computed from the Tauc relationship[27]

$$(\alpha h v)^{1/m} = A(h v - \text{E}_g), \tag{4}$$

here, $\alpha$ is the absorption coefficient, $hv$ is the photon energy [eV], $A$ is a dimensionless constant; $m$ = 1/2 ($m$ = 2) for direct (indirect) allowed electronic transition modes. TiO$_2$ is generally believed to be an indirect bandgap semiconductor, and its $m$ = 2.[28] The estimated E$_g$ decreases as plasma power increases (Figure 2c), which is consistent with our previous work.[18b]

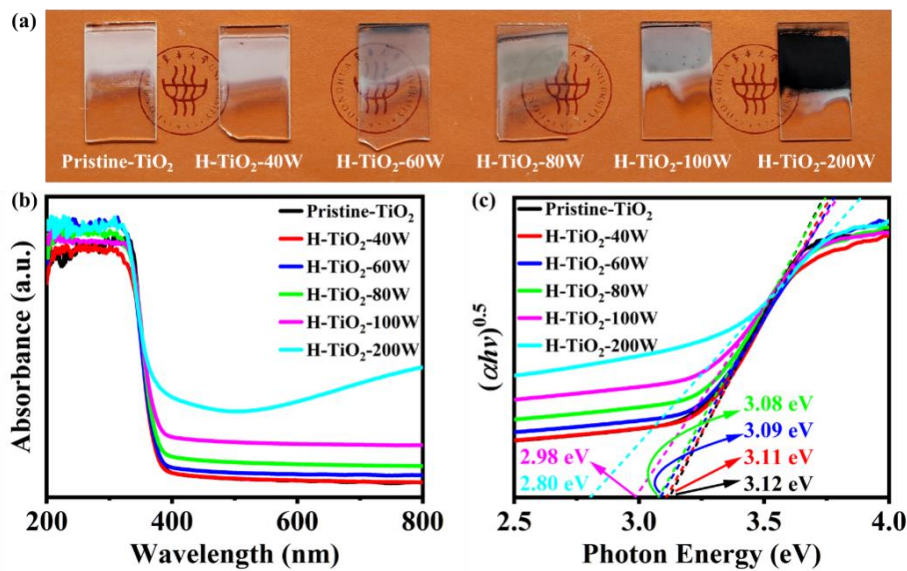

**Figure 2.** (a) Photographs, (b) UV-Vis absorption spectra, (c) Tauc plot of pristine and H–TiO$_2$ films.

### 2.3. Morphological characterization

Figure 3 shows top- and section-view SEM images of pristine and H–TiO$_2$ thin films. Dense morphologies, characterized by agglomerated polyhedral grains ranging from 0.1 to 1.5 μm in size, are observed in all samples. The cross-sectional SEM images reveal that the TiO$_2$ films consist of large and well-connected crystal columns, 0.1–0.5 μm in length. The bulk morphologies of hydrogenated TiO$_2$ are not very different from that of the pristine film; the RF plasma hydrogenation process had only a negligible effect on the film morphology. Note that a dense TiO$_2$ film is essential to efficiently transport photoexcited carriers (electron-hole pairs) between grains, which is how photodetection is carried out. Yan et al.[23] demonstrated that TiO$_2$ films (20 nm, grown on the quartz substrate via an atomic layer deposition system) after H$_2$ plasma treatment (ICP, 25.8–27.1 mTorr, 3000 W, 300 °C, 50 sccm H$_2$ flow rate, 30 min) exhibits a surprisingly high conductivity with the value of 72.9 S cm$^{-1}$. Kang et al.[29] reported that the conductivity of TiO$_2$ pellets after annealed under H$_2$ and Ar flow at 700 °C for 10 h ranged from $10^{-2}$ to $10^{-1}$ S cm$^{-1}$. Our result is consistent with theirs that a dense structure of H–TiO$_2$ films is beneficial for improving conductivity, but with much rapid (10 min), energy saving and efficient compared to others work.

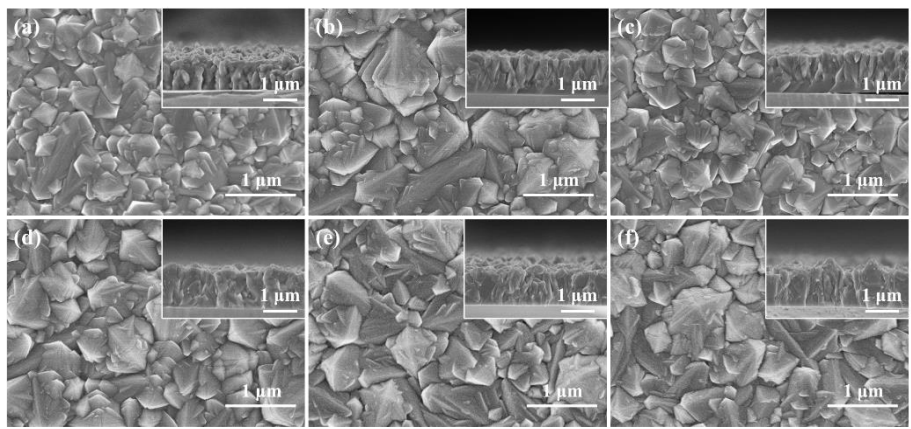

**Figure 3.** Top- and (inset) section-view SEM images of (a) pristine TiO₂; TiO₂ hydrogenated at (b) 40W, (c) 60 W, (d) 80 W, (e) 100 W, (f) 200 W.

## 2.4. Microstructure analysis

We analyzed the pristine and hydrogenated films by XRD (Figure 4a) and Raman (Figure 4b) spectroscopy. The well-defined diffraction peaks in the XRD spectra indicate that the films are largely composed of anatase TiO$_2$ (JCPDF#21-1272). A strong diffraction peak at 25.3° corresponding to (101) plane of anatase is observed in all samples, while a new peak located near 26.3° (marked with an arrow) only appears in H–TiO$_2$-200W; it corresponds to Ti$_3$O$_5$ (JCPDF#40-0806). This peak suggests that a phase transition occurs under sufficiently intense RF AP He/H$_2$ plasma treatment. Furthermore, the (101) peak intensity decreases with increasing plasma power, which is attributable to the introduction of defect states by the plasma. Both the phase transition and the reduction in peak intensity indicate an increase in the concentration of defects within TiO$_2$ crystals with increasing plasma power.

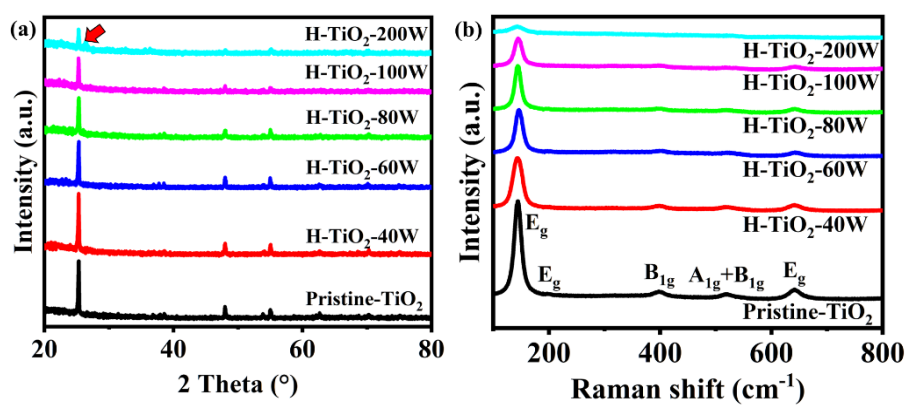

**Figure 4.** (a) XRD patterns and (b) Raman spectra of pristine and H–TiO$_2$ films.

We calculate values of various crystal parameters, including the average crystallite sizes ($D$), dislocation densities ($\delta$), micro-strains ($\varepsilon$) and interplanar spacing of the (101) plane ($d$), employing Equations 5 to 8 from the reference[30] and Figure S2.

$$D = \frac{k\lambda}{\beta \cos \theta}; \tag{5}$$

$$\delta = \frac{1}{D^2}; \tag{6}$$

$$\varepsilon = \frac{\beta}{4 \tan \theta}; \tag{7}$$

$$2d \sin \theta = \lambda. \tag{8}$$

Here, $k = 0.89$ is the Scherrer constant; $\lambda = 0.154056$ nm is the X-ray wavelength; $\theta$ is the diffraction angle of the Bragg peak; and $\beta$ is the full width at half maximum (FWHM) of the peak. We expect $D$ to lie between 1 and 100 nm in the anatase (101). As shown in Table 2, our films have $D$ between 39 and 47 nm, suggesting that our results are reliable. We observe that the crystallite sizes $D$ decrease with increasing plasma power, while $\delta$ and $\varepsilon$ increase. We thus attribute the broadening of the (101) diffraction peak to the simultaneous reduction of particle size and increasing presence of micro-strain within the microcrystals[30a], resulting from defects in the $TiO_2$ lattice. The values we compute for the anatase (101) interplanar spacing $d$ agree with previously reported data.[30a]

**Table 2.** Lattice parameters of pristine and H–$TiO_2$ films based on measured XRD data. All parameters refer to the anatase (101) crystal plane.

| Sample | β [°] | D [nm] | δ [$10^{-3}$ $nm^{-2}$] | ε | d [Å] |
|---|---|---|---|---|---|
| pristine $TiO_2$ | 0.174 | 46.20 | 0.47 | 3.39 | 3.52 |
| H–$TiO_2$-40W | 0.180 | 44.67 | 0.50 | 3.50 | 3.52 |
| H–$TiO_2$-60W | 0.187 | 42.96 | 0.54 | 3.64 | 3.52 |
| H–$TiO_2$-80W | 0.188 | 42.89 | 0.54 | 3.65 | 3.52 |
| H–$TiO_2$-100W | 0.192 | 41.90 | 0.57 | 3.74 | 3.52 |
| H–$TiO_2$-200W | 0.205 | 39.37 | 0.65 | 3.98 | 3.52 |

The typical Raman active modes of anatase are observed at 144 ($E_g$), 197 ($E_g$), 399 ($B_{1g}$), 513 ($A_{1g}$), 519 ($B_{1g}$), and 639 ($E_g$) $cm^{-1}$.[31] The intensity of all the peaks decreases with increasing hydrogenation in the H–$TiO_2$ films, indicating the disruption of $TiO_2$ crystal structure induced by the plasma. Furthermore, the $E_g$ peak around 144 $cm^{-1}$ broadens noticeably, which we attribute either to strain into the crystal lattice or to increase in the density of defects.[28]

## 2.5. Chemical composition analysis

Ti, O, C and N peaks are observed in XPS spectra (Figure S3); C and N are believed to originate from contamination during storage and testing. The Ti 2$p$ core-level spectrum of pristine TiO$_2$ has two prominent peaks at binding energies (BE) of 459.02 and 464.71 eV (Figure 5a). They correspond respectively to 2$p_{3/2}$ and 2$p_{1/2}$ peaks of Ti$^{4+}$.[32] This indicates that the titanium in pristine TiO$_2$ is in its standard oxidation state.[23, 33] The Ti spectrum of H–TiO$_2$-40W (Figure 5b) is similar to that of pristine TiO$_2$, indicating that Ti$^{4+}$ is not reduced in detectable amounts. However, the Ti spectrum of H–TiO$_2$-200W exhibits six discernible peaks (Figure 5c). Apart from the primary peaks associated with Ti$^{4+}$ (458.48 and 464.50 eV), additional shoulder peaks were observed at lower binding energies: 456.78, 463.08, 455.17, and 461.21 eV. These indicate the presence of Ti$^{3+}$ and Ti$^{2+}$ species.[19, 34] The ratio of (Ti$^{3+}$ + Ti$^{2+}$) to total titanium in H–TiO$_2$-200W is approximately 44% (Table 3). Additionally, a negative shift of the Ti 2p peaks is observed in H–TiO$_2$, which may be attributed to the formation of OV.

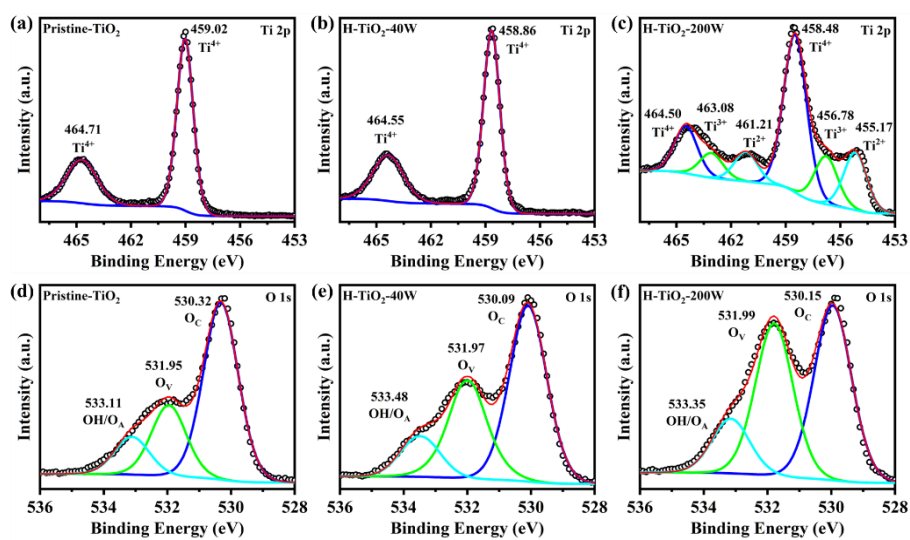

**Figure 5.** High-resolution (a–c) Ti 2p and (d–f) O 1s XPS of pristine TiO$_2$, H–TiO$_2$-40W and H–TiO$_2$-200W. Open circles and lines represent the experimental and the fitted data, respectively.

The O 1s core level spectra of pristine TiO$_2$, H–TiO$_2$-40W, and H–TiO$_2$-200W samples (Figure 5d–f) are similar; each can be deconvoluted into three distinct peaks. These are attributed to crystal lattice oxygen (O$_C$, Ti$^{4+}$–O); defect oxygen (O$_V$, Ti$^{3+}$–O and/or Ti$^{2+}$–O); and surface hydroxyl groups (Ti–OH bonds) or adsorbed oxygen (O$_A$).[33a, 34a] The ratio of defect oxygen to bulk oxygen, O$_V$ / (O$_V$ + O$_C$), is 28.57% (35.39%, 46.54%) for pristine TiO$_2$ (H–TiO$_2$-40W, H–TiO$_2$-200W). This trend indicates that the RF hydrogen plasma promotes the introduction of oxygen vacancies (equivalently, reduced Ti$^{3+}$/Ti$^{2+}$ species) into the TiO$_2$ lattice.

**Table 3.** Binding energy (BE) and percentage of each atom (at%) from high-resolution XPS Ti 2p and O 1s spectra for pristine TiO$_2$, H–TiO$_2$-40W, and H–TiO$_2$-200W

| Peaks | Pristine TiO$_2$ | | H–TiO$_2$-40W | | H–TiO$_2$-200W | |
|---|---|---|---|---|---|---|
| | BE [eV] | at% | BE [eV] | at% | BE [eV] | at% |
| Ti$^{4+}$ 2p$_{3/2}$ | 459.02 | 15.8 | 458.86 | 10.4 | 458.48 | 16.3 |
| Ti$^{3+}$ 2p$_{3/2}$ | - | - | - | - | 456.78 | 5.0 |
| Ti$^{2+}$ 2p$_{3/2}$ | - | - | - | - | 455.17 | 6.0 |
| Ti$^{4+}$ 2p$_{1/2}$ | 464.71 | 7.5 | 464.55 | 5.1 | 464.50 | 4.8 |
| Ti$^{3+}$ 2p$_{1/2}$ | - | - | - | - | 463.08 | 2.6 |
| Ti$^{2+}$ 2p$_{1/2}$ | - | - | - | - | 461.21 | 3.1 |
| O$_C$ | 530.32 | 22.0 | 530.09 | 24.1 | 530.15 | 11.6 |
| O$_V$ | 531.95 | 8.8 | 531.97 | 13.2 | 530.09 | 10.1 |
| OH/O$_A$ | 533.11 | 4.7 | 533.48 | 5.9 | 533.35 | 3.6 |

We also estimated the electron temperature ($T_e$) and gas temperature ($T_g$) of plasma operated at 200 W via OES spectra in Figure S4. $T_e$ and $T_g$ are 8500 K (0.73 eV) and 700 K, respectively. The bond energies of Ti-O bonds in TiO$_2$ crystal are 1.06 and 0.71 eV, respectively (details in Supporting Information). Obviously, the energy of high energy electrons in plasma is close to the bond energy of Ti-O bonds. Hence, He and H$_2$ molecules were split abundantly into charged active species due to the collision of high energy electrons. Numerous charged active species were accelerated by sheath electric field and bombard the TiO$_2$ films surface with sufficient energy[18b]. It resulted in the dissociation of Ti–O bonds and desorption of O atoms[35], creating OVs, hydroxyl groups (Ti–OH bonds), Ti$^{3+}$ and Ti$^{2+}$ species in the region of films surface. The presence of these defect states leads to the tailoring of the band gap of H-TiO$_2$, which is beneficial for the improvement of visible light absorption and enhanced conductivity.

## 2.6. Electronic structure analysis

The spin-up and spin-down electronic states of TiO$_2$ are degenerate, implying that stoichiometric TiO$_2$ is a non-magnetic semiconductor (Figure 6a). When doped with oxygen vacancies, however, the degeneracy is lifted for states in the conduction band (CB), particularly near the conduction band minimum (CBM) (Figure 6d, g). This suggests that OV doping induces magnetism, supporting the conclusion of previous studies.[14, 36] OV doping also raises the Fermi level to the conduction band. The CBM and valence band maxima (VBM) for $TiO_2$, $TiO_{1.9375}$ and $TiO_{1.875}$ are located at different wave vectors (Figure S5), indicating that both pristine and hydrogenated anatase $TiO_2$ are indirect semiconductors.

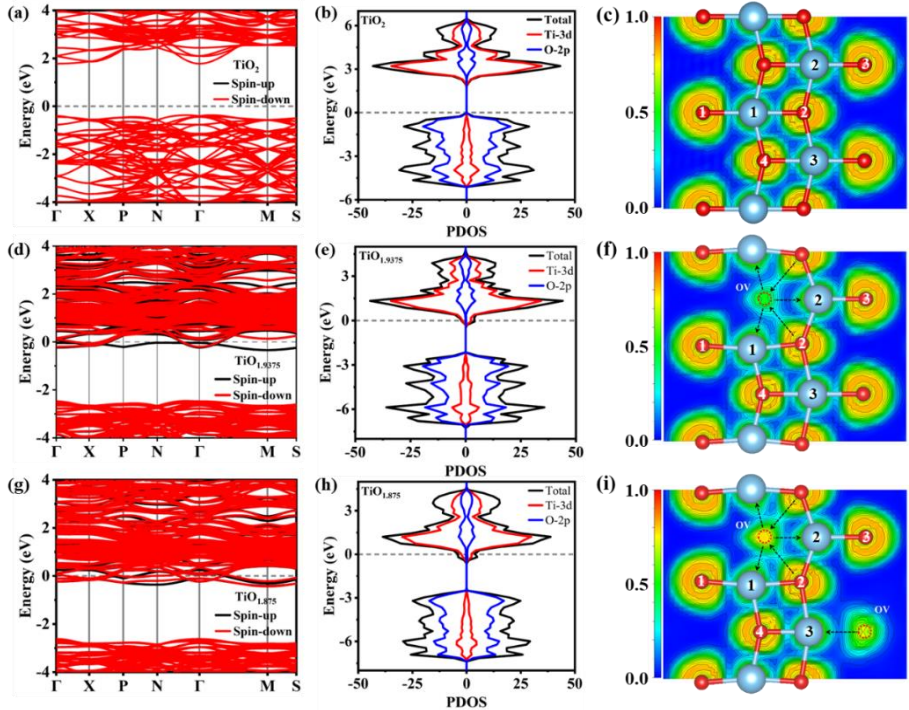

**Figure 6.** Electronic band structures, DOS, and ELF cross-section of (010) plane with overlaid ball-and-stick diagrams: (a–c) $TiO_2$, (d–f) $TiO_{1.9375}$, (g–i) $TiO_{1.875}$. Atoms near oxygen vacancies are labeled (1), (2), (3) and (4).

The valence bands near the Fermi level consist primarily of 2$p$ electrons from oxygen, with smaller contribution from Ti 3$d$ electrons (Figure 6b, e, h). The VBM is determined by the O 2$p$ orbital electrons. On the other hand, the conduction bands near the Fermi level are predominantly composed of Ti 3$d$ orbital electrons, with a smaller contribution from the O 2$p$ electrons. The CBM is thus determined by the Ti 3$d$ orbital electrons.

The results obtained from the projected density of states (PDOS) analysis are consistent with the band structures. The PDOS of pristine $TiO_2$ is symmetric in spin (Figure 6b), while those of $TiO_{1.9375}$ and $TiO_{1.875}$ are not (Figures 6e, h). This asymmetry further suggests magnetic properties in $TiO_{2-x}$.[14, 36] Increasing the OV content decreases the PDOS near the VBM slightly, while increasing it significantly near the CBM relative to that of $TiO_2$. The Fermi level (set to 0 in Figure 6b, e, h) shifts from the VBM in pristine $TiO_2$ to the CB in $TiO_{1.9375}$ and $TiO_{1.875}$. We explain the Fermi shift as follows: each OV generates two surplus electrons in the crystal, and they redistribute into the three Ti 3$d$ orbitals nearest to the OV.[37] These redistributed Ti 3$d$ states reside primarily in the CBM and are the primary cause for the transition from $Ti^{4+}$ to $Ti^{3+}$ species in the lattice. The excess electrons also shift the Fermi level up, strengthening the *n*-type semiconductor character of OV-doped $TiO_2$.[37] Note that, despite the band structure of OV-doped $TiO_2$ resembling that of a conductor, the conductivity of OV-doped $TiO_2$ is still relatively small due to the small DOS at the occupied CBM. Thus, $TiO_2$ with OV is still classified as a *n*-type semiconductor with some metallicity, rather than a true conductor.

We also plot the electron localization function (ELF) for the (010) plane (Figure S6, Figure 6c, f, i). The ELF value at a specific position is the probability of finding an electron at that location, accounting for the presence of neighboring electrons.[38] The ELF ranges from 0 to 1; values close to 1 indicate a high probability of electron localization, while 0 suggests complete delocalization or the absence of electrons. An ELF value close to 1/2 characterizes electron gas-like behavior.[38] Thus, the ELF visualizes the electrons' distribution within the system (Figure S6). The ELF of Ti atoms ranges from 0.5 to 1, indicating a locally high electron density and the presence of a surrounding electron cloud. The electrons surrounding O atoms are more delocalized, resulting in a lower electron density. The ELF value between Ti and O is close to 1/2, as expected for covalent Ti–O bonds. Introducing OV results changes the lattice constants (< 2.634%) and cell volume (< 0.343%) slightly (Table S1), but the local structure of the lattice near the OV is altered significantly (Figures 6f, i). The electron density around the atoms near the defect tends to occupy the empty sites.[39] Consistent with previous studies, and with our XRD and Raman analyses, we observe that the O cores exhibit slight displacements toward the OV, while the Ti cores move away (Table S3).[39]

## 2.7. Photo-response characteristics and smart feature analysis

The "5S" criterion—sensitivity, signal-to-noise ratio, speed, selectivity, and stability—is widely accepted as the standard for evaluating the optoelectronic performance of UV PDs.[1d] Responsivity, *R*, holds significant importance in assessing the sensitivity of a photoelectric device. It can be calculated as[9a]

$$R = \frac{I_{ph}}{PA} = \frac{I_L - I_D}{PA}, \tag{9}$$

where $I_{ph}$ is the net photocurrent, *P* is the incident power density (or irradiation intensity), *A* is the effective irradiated area, and $I_L$ ($I_D$) is the current under UV illumination (in the dark).

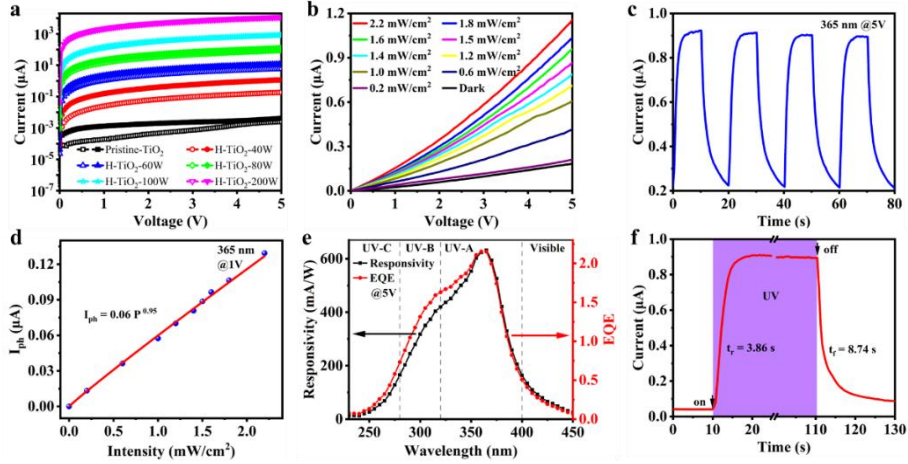

**Figure 7.** (a) Typical *I-V* curves of H–TiO$_2$ films measured in the dark and under 2.2 mW cm$^{-2}$ UV irradiation. Hollow dots: $I_D$, solid dots: $I_L$. (b) Typical *I-V* curves of H–TiO$_2$-40W under UV irradiation at various voltages. (c) Optical switching characteristics of H–TiO$_2$-40W at 5 V under 2.2 mW cm$^{-2}$ UV irradiation. (d) Variation in $I_{ph}$ with UV intensity. (e) Typical photoresponsivity spectrum and EQE as a function of wavelength at 5 V. (f) Typical transient photo-response curves of H–TiO$_2$-40W under 2.2 mW cm$^{-2}$ 365 nm UV illumination.

Figure 7a presents a comparison of $I_L$ (solid dots) and $I_D$ (hollow dots) for pristine and H–TiO$_2$ films. H–TiO$_2$-200W has the highest $I_L$ (11.81 mA at 5 V). However, it also has the highest $I_D$ (11.76 mA at 5V, almost as high as $I_L$), which we attribute to its high conductivity (10.2 S cm$^{-1}$). Like the film hydrogenated at 200 W, H–TiO$_2$-100W and H–TiO$_2$-80W exhibit only small differences in $I_L$ and $I_D$. The difference between $I_L$ and $I_D$ for H–TiO$_2$-60W (about 11.81 μA, 5.88 μA at 5V), H–TiO$_2$-40W (about 1.16 μA, 0.18 μA at 5V), and pristine TiO$_2$ (about 4.10 nA, 2.55 nA at 5V) are significant, however. $I_D$ is considered background noise for PDs, so a high $I_D$ is undesirable. Advanced PDs should have low $I_D$ and high $I_L$ to minimize background noise and maximize overall performance. Table 4 lists the signal-to-noise ratios ($I_L - I_D$) / $I_D$ values for all samples.

**Table 4.** ($I_L - I_D$) / $I_D$ for pristine and H–TiO$_2$ films.

| Sample | $I_D$ [μA] | $I_L$ [μA] | ($I_L - I_D$) / $I_D$ |
| --- | --- | --- | --- |
| H–TiO$_2$-200W | 11.76×10$^3$ | 11.81×10$^3$ | 0.004 |
| H–TiO$_2$-100W | 818.03 | 867.91 | 0.061 |
| H–TiO$_2$-80W | 93.11 | 120.40 | 0.293 |
| H–TiO$_2$-60W | 5.88 | 11.81 | 1.009 |
| H–TiO$_2$-40W | 0.18 | 1.16 | 5.444 |
| Pristine TiO$_2$ | 2.55×10$^{-3}$ | 4.10×10$^{-3}$ | 0.608 |

From Table 4, it is found that the signal-to-noise ratios ($I_L - I_D$) / $I_D$ values of H–TiO$_2$-200 W is very low. Pristine TiO$_2$ shows a discernable signal-to-noise ratio at low voltage, but it is not stable at greater voltage and its $I_L$ is very low. The signal-to-noise ratios ($I_L - I_D$) / $I_D$ values of 40W is the best, so we select it as the optimal film.

In H–TiO$_2$-40W, $I_L$ increases with UV intensity from 0.2 to 2.2 mW cm$^{-2}$, as well as with the applied bias voltage (Figure 7b). A distinctive periodic current is also observed under intermittent UV illumination (Figure 7c), indicating a reversible switching between states of low (0.22 mA) and high (0.91 mA) resistivity. This switching may be related to the fact that the carrier concentration is positively correlated to the absorbed photon flux;[40] typically, the relationship between $I_{ph}$ and UV intensity ($P$) follows a power law.[41] We fit this (Figure 7d) as

$$I_{ph} = 0.06P^{0.95}, \qquad (10)$$

since the calculated exponent (0.95) is close to 1, we apply a linear relation for calibration across a wide range of UV intensity, 0–2.2 mW cm$^{-2}$; this is a standard practice.[3a] The repeatability of the switching and the linearity of the response demonstrate that H–TiO$_2$-40W can accurately monitor and report UV intensity and is therefore a good choice for a UV PD. We show the responsivity spectrum and external quantum efficiency (EQE) of H–TiO$_2$-40W in Figure 7e. The EQE reflects the energy conversion efficiency from light to electricity, and is calculated as

$$\text{EQE} = R\frac{hv}{e} \times 100\%, \qquad (11)$$

where $hv$ denotes the energy of a single photon and $e$ is the elementary charge. Both the responsivity and EQE increase with increasing wavelength, then decrease as the wavelength increases further. The highest responsivity (632.35 mA W$^{-1}$) and EQE (2.15) are observed around 365 nm.

The response speed is commonly assessed by measuring the response time, defined by the time it takes for the response photocurrent to rise from 10% to 90% of its maximum value (rising time, $t_r$) and to fall from 90% to 10% (falling time, $t_f$).[1d] For H–TiO$_2$-40W at 5 V bias voltage under irradiation by 2.2 mW cm$^{-2}$ of 365 nm UV, $t_r$ ($t_f$) is 3.86 s (8.74 s) (Figure 7f). Compared to other UV PDs, which have response times of milliseconds[9b, 42] or even microseconds,[40, 43] our UV PD is slow and is not unexpected for TiO$_2$-based photodetectors.[44] Because the response speed is influenced by various additional factors, including fine details of the device design and manufacturing processes; the presence of defect or trap states within the semiconductor;[1d] and the oxygen adsorption and desorption

WILEY-VCHprocesses on their surfaces,[9a] interface engineering and trap state modulation are effective strategies for regulating the response speed of TiO$_2$ based UV PDs.[45] Further research and technological innovation are necessary in order to improve the response speed of H-TiO$_2$ based UV PDs, which will unlock their full potential for sensing, imaging, and communication systems. However, our samples sensitivity is excellent. The lowest UV intensity detectable is less than 20 μW cm$^{-2}$, limited by measurement range (Figure S7).

**2.8. Application in UV exposure measurement**

The photo-response test demonstrated that the H–TiO$_2$-40W film is suitable for wearable UV monitoring. We designed an interface circuit and incorporated it with the film to report UV intensity. Figure 8a shows the UV PD being worn outdoors. Ambient UV intensity is measured in real time and transferred to a smartphone when the PD is worn on the wrist and exposed to UV radiation. UV intensity is converted into an electrical signal ($I_{ph}$) by the H–TiO$_2$-40W film. This signal is converted to a digital representation by the interface circuit and sent to the smartphone via Wi-Fi (Figure 8b). Our photodetector is an efficient, pragmatic, and convenient approach for notifying people about the potential risks of excessive UV exposure. In Figure 8c, we compare our UV PD and a commercial UV radiometer. UV intensity in this test was adjusted by controlling the distance between a UV sterilization lamp and the device. The result indicates almost perfect agreement (up to linear calibration) from 0–40 W m$^{-2}$ of incident UV, confirming that our PD is of comparable effectiveness and accuracy to commercial UV radiometers. In Figure 8d, we present daily variation UV intensity, Global Solar UV Index (UVI) and exposure categories for three typical weather types (sunny, cloudy and rainy) during summer 2023 in Shanghai. (More details about UVI and exposure categories can be found in the supplementary material.) The UV intensity increases in the morning (05:00 to 12:00) and decreases over the afternoon (12:00 to 19:00). However, the UV intensity is quite different under different weather types. On sunny days, UV intensity is strongest and most stable, distributed roughly symmetrically around noon. On the cloudy day, the fluctuation of UV intensity is relatively large, decreasing dramatically in the shadow of a cloud. On rainy days, UV intensity is reduced even further. When ambient ultraviolet intensity is sufficiently high, it is essential to take appropriate precautions (see Table S4).

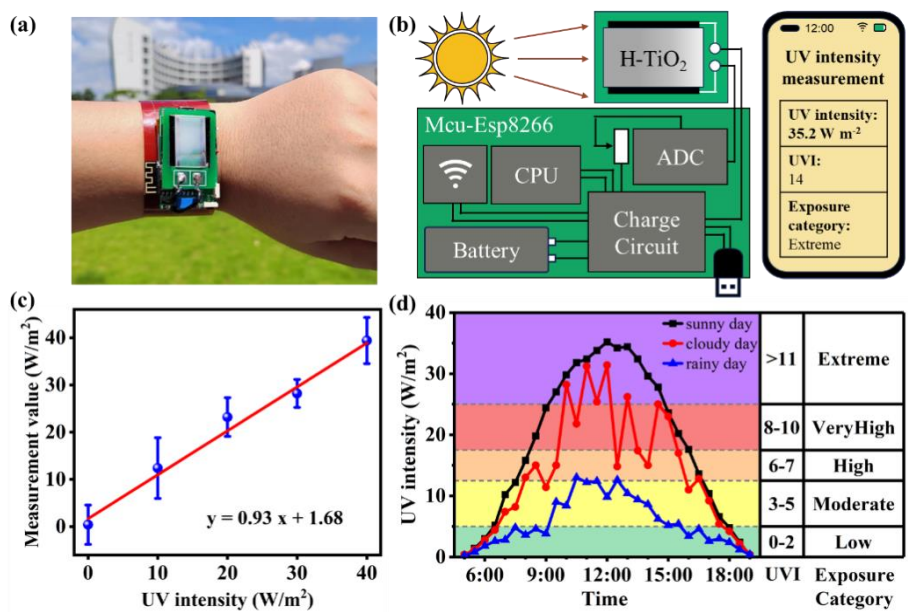

**Figure 8.** (a) Photograph of wearable UV PD in ambient environment, (b) Schematic diagram of the PD, (c) Comparing our PD and a commercial radiometer, (d) Daily variation of UV intensity under different weather conditions in the Shanghai summer.

## 3. Conclusion

We developed a real-time and wearable UV PD based on an H–TiO$_2$ film, which we synthesized with radio frequency He/H$_2$ plasma at atmospheric pressure. One sample in particular, H–TiO$_2$-200W, demonstrated a remarkably high conductivity of 10.2 S cm$^{-1}$, which is 9 orders of magnitude higher than that of pristine TiO$_2$. The visible light absorption of the H–TiO$_2$ films increase with the plasma treatment power, implying a corresponding decrease in the bandgap. SEM images demonstrate that pristine and H–TiO$_2$ films are composed of grains ranging in size from 0.1 to 1.5 μm. These grains are densely packed and interconnected, enabling the efficient transmission of photoexcited carriers between them. XRD patterns and Raman spectra indicate that the plasma damages the TiO$_2$ crystal structure; XPS results demonstrate that oxygen vacancies have been introduced into the lattice, creating Ti$^{3+}$ and Ti$^{2+}$ species. DFT calculations confirm that the OVs generate additional electrons, shifting the Fermi level upward: TiO$_{2-x}$ displays some metallic character. The OVs also cause lattice distortion, which disrupts the long-range order.

H–TiO$_2$-40W shows almost 3 orders higher illumination current and 1 order higher ratio of ($I_L$ – $I_D$) / $I_D$ than that of pristine TiO$_2$. It has a responsivity of 632.35 mA W$^{-1}$ and an EQE of 2.15 (5 V bias, 365 nm UV illumination), which is the optimal candidate for our UV PD. H–TiO$_2$-40W also shows a linear relationship between $I_{ph}$ and UV intensity, which is crucial for its function as a photodetector. It displays a wide spectral response range, covering both UV-A and UV-B. However, its $t_r$ ($t_f$) is 3.86 s (8.74 s); improving the response time is an open challenge.

We fabricated a sensitive wearable, real-time UV PD by combining H–TiO$_2$-40W and an interface circuit, using it to monitor daily variations in UV. The UV intensity, UVI, and exposure categories under three typical weather conditions (sunny, cloudy and rainy) were measured, and some UV protection measures were also provided. Our UV PD integrates with commercial data collectors and analyzers easily due to the installation of Wi-Fi. This work paves a new way for wide-bandgap semiconductor materials to be used in UV photodetectors through facile, rapid, and low-cost plasma technology.

## 4. Experiments and Methods

*Preparation of H–TiO$_2$ film*: Pristine white TiO$_2$ films were deposited on a quartz substrate by RF AP pulse-modulated PECVD (120 W input power, 100 ms on time, 50% duty cycle),[46] employing a 13.56 MHz plasma generator (RF-10S/PWT) and a square quartz chamber (1 mm thick, 2 mm gap). The flow rates of helium (He), oxygen (O$_2$) and titanium tetrachloride (TiCl$_4$, carried by He) were set as 600, 3 and 5 sccm, respectively. A white pristine TiO$_2$ film was obtained on the substrate after 30 min.

Then, the pristine TiO$_2$ film was treated by RF AP He/H$_2$ plasma for 10 min to prepare the hydrogenated H–TiO$_2$ film.[18b] The RF plasma generator (RF-10S/PWT) was operated in continuous mode (duty cycle = 100%), with treatment power varied from 40 to 200 W. He (H$_2$) was used as the discharge gas (reducing agent), with a 600 sccm (5 sccm) flow rate. The samples were named pristine TiO$_2$, H–TiO$_2$-40W, H–TiO$_2$-60W, H–TiO$_2$-80W, H–TiO$_2$-100W and H–TiO$_2$-200W, according to the plasma treatment power.

*Device design and assembly*: An interface circuit is an essential component of UV PD. It consists primarily of an analog-to-digital converter (ADC), microcontroller (MCU), Wi-Fi module, battery, and charge circuit. These commercial electronic components were carefully assembled and soldered on a printed circuit board (PCB). The PCB (5.0 × 2.7 cm$^2$) was designed by Altium Designer software. The H–TiO$_2$ film was integrated with the interface circuit; its real-time photocurrent signals are converted into digital signals by ADC. Then, these digital signals are encoded by the MCU and sent to a Python script[47] via the Wi-Fi module. Finally, the received data are processed and analyzed to obtain the UV values. For convenient user access to these results, a web page was developed for displaying these values. It can be accessed from its IP address on a mobile browser.

*Characterizations*: The morphologies of the films were studied using a scanning electron microscope (SEM, Hitachi S-4800). Their crystal structures were characterized using X-ray diffraction (XRD, D/max-2550+/PC) and Raman spectra (inVia-Reflex). The surface composition and chemical states were measured with an X-ray photoelectron spectrometer (XPS, Thermo Escalab 250Xi). The optical absorbance spectra were obtained by a UV-vis spectrophotometer (Shimadzu UV-2600). The conductivities were measured by a four-point probe measurement system (RTS-9). Hall coefficients ($R_H$), carrier concentrations ($n$), and mobilities ($\mu$) were determined by a Hall measurement system (Ecopia HMS-7000) with a magnetic field of 0.5 T at room temperature. The current-voltage (*I-V*), current-time (*I–t*) and photo-response characteristics were investigated by a semiconductor characteristic analyzer system (Keithley 4200A) at room temperature. The intensity of UV irradiation was measured using a commercial UV radiometer (HANDY UV-A).

*Theoretical calculations*: All theoretical investigation of these films were conducted with density functional theory (DFT) in the Vienna *ab initio* simulation package (VASP).[48] We use the projector-augmented wave (PAW)[49] method with the Perdew–Burke–Ernzerhof (PBE)[50] exchange-correlation functional. The plane-wave energy cutoff was set as 450 eV, the energy convergence criterion was placed as at $10^{-5}$ eV atom$^{-1}$, and the force convergence accuracy was set to 0.01 eV Å$^{-1}$. The k-mesh in the first Brillouin zone (BZ) were sampled by the Monkhorst–Pack method with a grid mesh of $7 \times 7 \times 7$.

Anatase $TiO_2$ belongs to the $I4_1/amd$ space group, with lattice parameters a = b = 3.78 Å, c = 9.62 Å, and α = β = γ = 90º. Consistent with prior studies and as shown in Figure S1, a $2 \times 2 \times 1$ $TiO_2$ supercell was selected to calculate the correlation properties and consider different concentrations of OV ($TiO_{2-x}$, x = 0, 0.0625, 0.125). The lattice parameters, cell volume and total energy of relaxed $TiO_{2-x}$ are shown in Table S1.

**Supporting Information**

Supporting Information is available from the Wiley Online Library or from the author.

**Acknowledgements**

This work was supported by the National Natural Science Foundation of China (12075054 and 12205040).

**Conflict of Interest**

The authors declare no conflict of interest.

**Data Availability Statement**

The data that support the findings of this study are available from the corresponding author upon reasonable request.

Highly conductive TiO$_2$ films are directly prepared by radio frequency atmospheric pressure plasma through facil and environmental efficient processes with short time, low temperature and open atmosphere. The films shows remakbly higher current and sensitivity under UV illumination. The optimum H-TiO$_2$ film is applied to UV radiation photodetectors and integrated with an interface circuit. This work paves a new way for the application of wide-bandgap semiconductor materials in UV photodetectors through facile, rapid, and low-cost plasma technology.

Y. Zhang, H. Wang, J. Cui, T. He, G. Qiu, Y. Xu*, and J. Zhang*

**An ultraviolet photodetector based on conductive hydrogenated TiO$_2$ film prepared by radio frequency atmospheric pressure plasma**

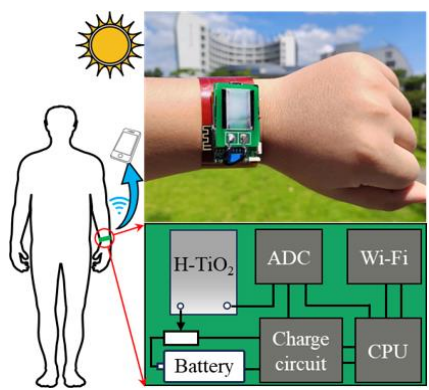

ToC figure

Supporting Information

# An ultraviolet photodetector based on conductive hydrogenated $TiO_2$ film prepared by radio frequency atmospheric pressure plasma

*Yu Zhang, Haozhe Wang, Jie Cui, Tao He, Gaote Qiu, Yu Xu\*, and Jing Zhang\**

Y. Zhang, H. Wang, J. Cui, T. He, G. Qiu
College of science, Donghua University, Shanghai 201620, China.
Y. Xu, J. Zhang
College of science, Donghua University, Shanghai 201620, China; Textile Key Laboratory for Advanced Plasma Technology and Application, China National Textile & Apparel Council, Shanghai 201620, China; Magnetic Confinement Fusion Research Center, Ministry of Education, Shanghai 201620, China.
Email: yuxu@dhu.edu.cn (Y.X.) and jingzh@dhu.edu.cn (J.Z.)

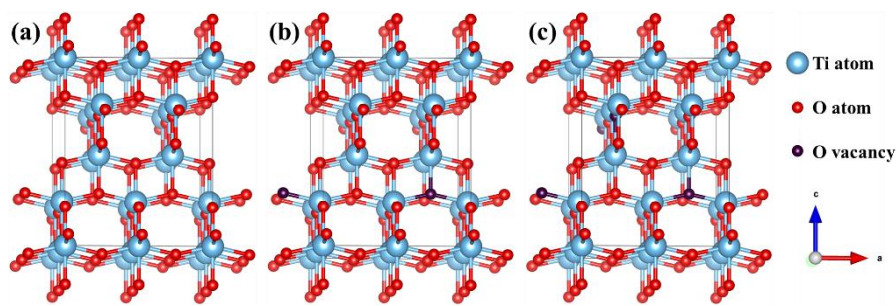

**Figure S1.** TiO$_{2x}$ (x=0, 0.0625, 0.125) supercell models with different molar number OV by removing some (0, 1, 2) oxygen atoms in a 2×2×1 supercell.

**Table S1.** Reduced lattice parameters, cell volume and total energy of TiO$_{2-x}$ (x=0, 0.0625, 0.125) after structural optimization

| Supercell | a, b, c [Å] | V [Å$^3$] | E [eV] | N [10$^{21}$ cm$^{-3}$] |
|---|---|---|---|---|
| TiO$_2$ | a = b = 3.792, c = 9.714 [This work] | 139.706 | -431.020 | 0 |
| | a = b = 3.804, c = 9.761 [1] | | | |
| | a = b = 3.798, c = 9.735 [2] | | | |
| | a = b = 3.79, c = 9.80 [3] | | | |
| TiO$_{1.9375}$ | a = 3.802, b = 3.807, c = 9.649 [This work] | 139.672 | -421.244 | 1.790 |
| | a = b = 3.790, c = 9.910 [1] | | | |
| | a = b = 3.79, c = 9.81 [3] | | | |
| TiO$_{1.875}$ | a = 3.784, b = 3.798, c = 9.689 [This work] | 139.227 | -411.419 | 3.591 |

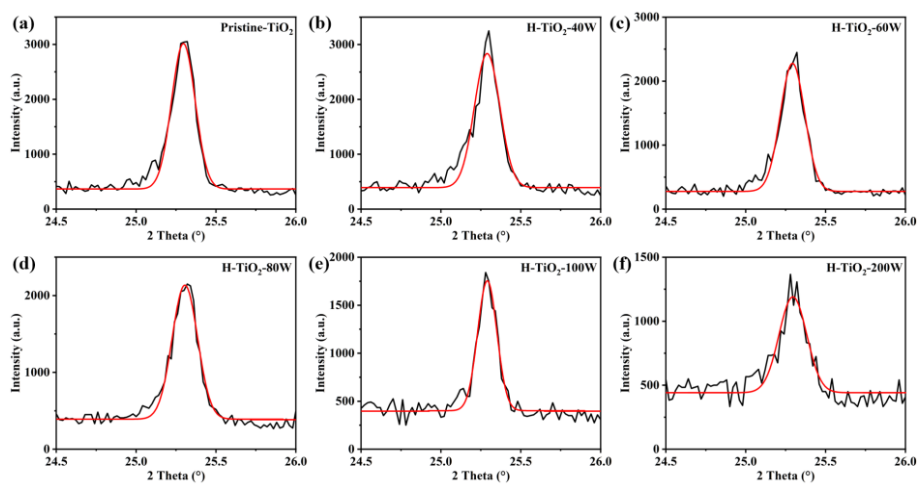

**Figure S2.** XRD patterns associated with anatase (101) peak.

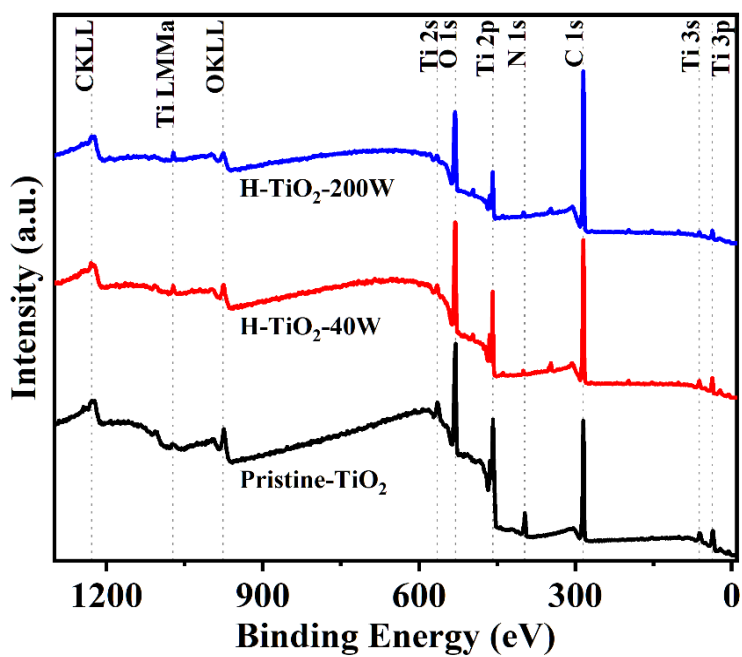

**Figure S3.** XPS full survey scan spectra of pristine- and H-TiO$_2$ films sample.

The electronic temperature ($T_e$) and gas temperature ($T_g$) of plasma operated at 200 W was estimated by Equation S1[4],

$$\ln\left(\frac{I\lambda}{gA}\right) = -\frac{E}{T_e k_B} + \text{constant} \quad \text{S1}$$

The gas temperature ($T_g$) of reaction chamber was estimated using MTO software (Measuring Temperature by OES, a spectral simulation program written in python 3.7)[5]. At atmospheric pressure, the OH radical's rotational temperature ($T_r$) could be accepted as an estimate of the gas temperature in plsama.

The wavelength of He lines obtained from the plasma OES result and compared with the data in the NIST (National Institute of Standards and Technology). Transitions between energy levels corresponding to wavelength, Einstein coefficient (A), energy values (E) and statistical weights (g) were obtained from NIST database to use data of optical emission spectrum of plasma. In addition, intensity values (*I*) were obtained from optical emission spectrum (Table 1).

Table S2 Intensity values, transitions, Einstein coefficients, energy values in transitions and statistical weights corresponding to each wavelength of the He spectral line

| λ (nm) | *I* (a.u.) | Transitions | A (×10⁸ s⁻¹) | $g_m - g_n$ | $E_m - E_n$ (eV) |
|---|---|---|---|---|---|

| | | Lower level–upper level | | | |
|---|---|---|---|---|---|
| 388.9 | **11293** | $2^3S$–$3^3P$ | 0.94746 | 3−3 | 19.8196−23.0070 |
| 587.5 | 109428 | $2^3P$–$3^3D$ | 7.0708 | 5−7 | 20.9642−23.0737 |
| 667.8 | 56820 | $2^1P$–$3^1D$ | 6.3705 | 3−5 | 21.2180−23.0741 |
| 706.5 | 10418 | $2^3P$–$3^3S$ | 1.5474 | 5−3 | 20.9642−22.7185 |
| 728.1 | 4159 | $2^1P$–$3^1S$ | 1.8299 | 3−1 | 21.2180−22.9203 |

Figure S4 presents the results of electron temperature ($T_e$) and gas temperature ($T_g$) of plasma operated at 200 W. $T_e$ and $T_g$ are estimated to be 8500 K (0.73 eV) and 700 K, respectively.

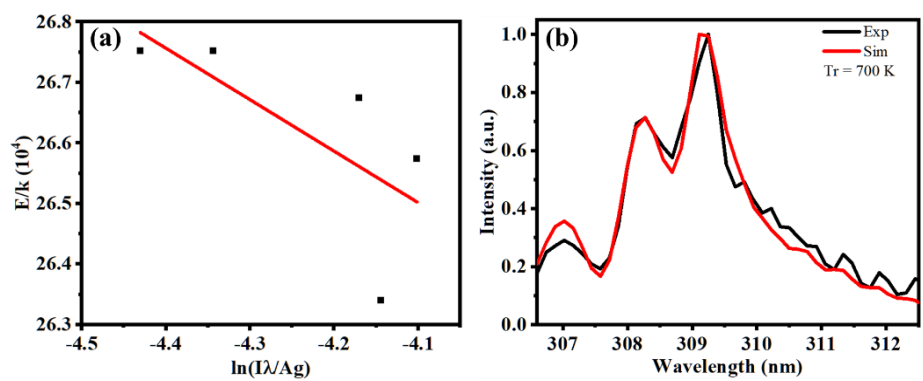

Figure S4. (a)The Boltzmann plot for the excitation temperature of plasma, (b) the experimental and simulation results of plasma gas temperature from OES results.

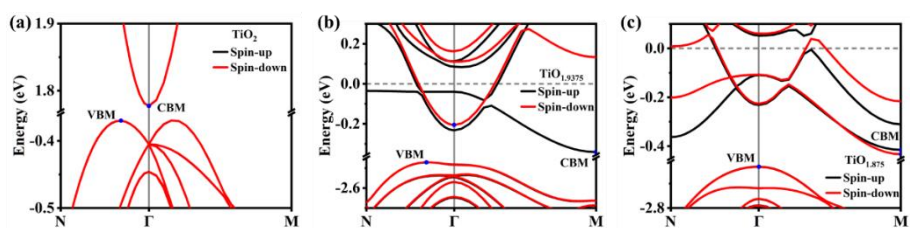

**Figure S5.** The band near the Fermi level of (a) $TiO_2$, (b) $TiO_{1.9375}$ and (c) $TiO_{1.875}$.

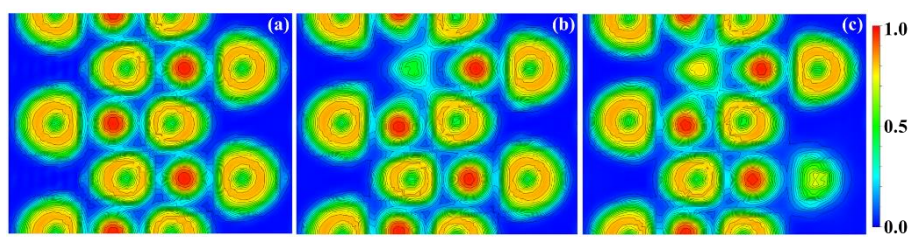

**Figure S6.** ELF cross-section of (0 1 0) plane of (a) $TiO_2$, (b) $TiO_{1.9375}$ and (c) $TiO_{1.875}$.

**Table S3.** List of bond length Ti-O near the OV.

| Bond | Bond length [Å] | | |
|---|---|---|---|
| | $TiO_2$ | $TiO_{1.9375}$ | $TiO_{1.875}$ |
| $Ti_1$-$O_1$ | 2.00218 | 1.97625 | 1.97051 |
| $Ti_1$-$O_2$ | 2.00218 | 1.98871 | 1.98041 |
| $Ti_1$-$O_4$ | 1.94348 | 1.85605 | 1.90239 |
| $Ti_2$-$O_2$ | 2.00218 | 1.90967 | 1.89454 |
| $Ti_2$-$O_3$ | 1.94348 | 1.89162 | 1.94322 |
| $Ti_3$-$O_2$ | 1.94348 | 2.05701 | 2.01621 |
| $Ti_3$-$O_4$ | 2.00218 | 1.99306 | 1.91789 |

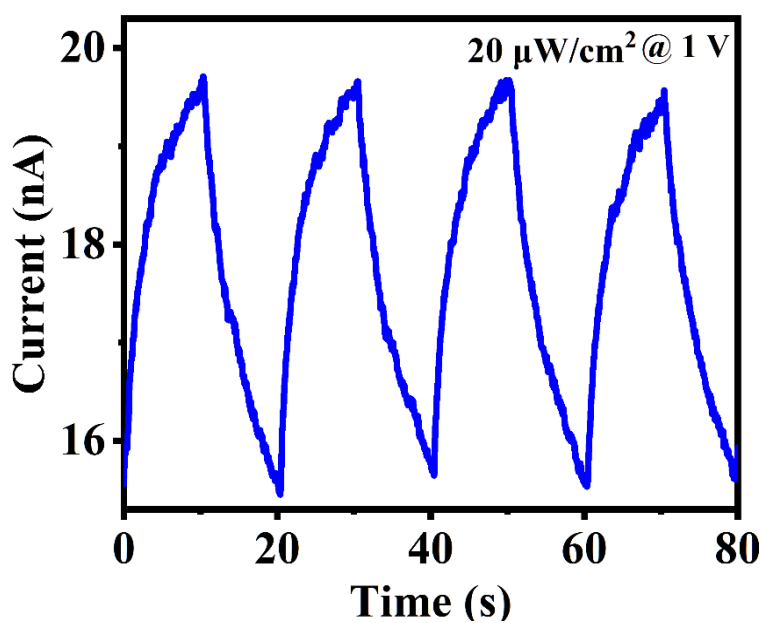

**Figure S7.** The transient photoresponse of H-$TiO_2$-40W under switched UV light with intensity of 20 μW cm$^{-2}$ at 1V.

In daily life, the level of UV radiation is assessed using a simpler numerical value known as the Global Solar UV Index (UVI) instead of UV intensity. This index facilitates a better understanding of the potential health risks and enables the general public to takes appropriate protection based on UVI value. UVI was developed by the World Meteorological Organization (WMO) in July 1994 and can be expressed as Equation S1 [6].

$$\mathrm{UVI} = k_{\mathrm{er}} \cdot \int_{250\mathrm{nm}}^{400\mathrm{nm}} E_\lambda \cdot S_{\mathrm{er}}(\lambda) \, \mathrm{d}\lambda, \qquad (S2)$$

WILEY-VCHwhere $E_\lambda$ is the solar spectral irradiance expressed in W m$^{-2}$ nm$^{-1}$ at wavelength $\lambda$ and d$\lambda$ is the wavelength interval used in the summation. $S_{er}(\lambda)$ is the erythema reference action spectrum (Equation S4), and $k_{er}$ is a constant equal to 40 m$^2$ W$^{-1}$. Depressingly, our PD cannot accurately measure the erythema-weighted UV spectrum. Fortunately, an alternative formula (Equation S2) for assessing the UVI was found in the "Forecasting Method for UV Index" (GB/T 36774-2018).

$$\text{UVI} \approx \frac{Q_{UV} \cdot C_{er}}{\Delta I}, \tag{S3}$$

where $Q_{UV}$ represents UV irradiance in W m$^{-2}$, corresponding to the UV intensity mentioned earlier. $C_{er}$ is the equivalent erythema correction factor, equal to 0.01. $\Delta I$ signifies the UV irradiance equivalent to the unit UVI, equal to 0.025 W m$^{-2}$. Thus, UVI can be simplified into Equation S3.

$$\text{UVI} = \frac{\text{UV intensity}}{2.5}, \tag{S4}$$

Generally, UVI is divided into 5 levels, as shown in Table S3.

**Table S4.** UVI classification standards.

| Exposure Category | UVI | UV Radiation [W m$^{-2}$] | Skin Redness Time [min] | Recommended Protection |
|---|---|---|---|---|
| Low | 0 - 2 | 0 - 5.0 | Negligible or none | Without additional protection |
| Moderate | 3 - 5 | 5.0 - 12.5 | 45 - 60 | Avoid prolonged exposure during intense sunlight and use sunscreen. |
| High | 6 - 7 | 12.5 - 17.5 | 30 - 45 | Avoid prolonged exposure during midday, use SPF 15 sunscreen, wear a hat, sunglasses and a sun umbrella etc. |
| Very High | 8 - 10 | 17.5 - 25.0 | 15 - 30 | Minimize outdoor activities, use SPF 30 sunscreen, wear a hat, sun umbralle, and sunglasses with UV protection, etc. |
| Extreme | 11$^+$ | > 25.0 | < 15 | Avoid outdoor activities, protect fully when outdoors, wear clothing and use sun umbrella with UV protection, etc. |

$S_{er}(\lambda)$ in 250 – 400 nm can be expressed as Equation S4.

$$S_{\text{er}}(\lambda) = \begin{cases} 1.0, & 250 \leq \lambda \leq 298 \text{ nm}; \\ 10^{0.094\,(298-\lambda)}, & 298 < \lambda < 328 \text{ nm}; \\ 10^{0.015\,(140-\lambda)}, & 328 \leq \lambda \leq 400 \text{ nm}; \end{cases} \quad (S5)$$

Figure S7 shows the plot of $S_{\text{er}}(\lambda)$.

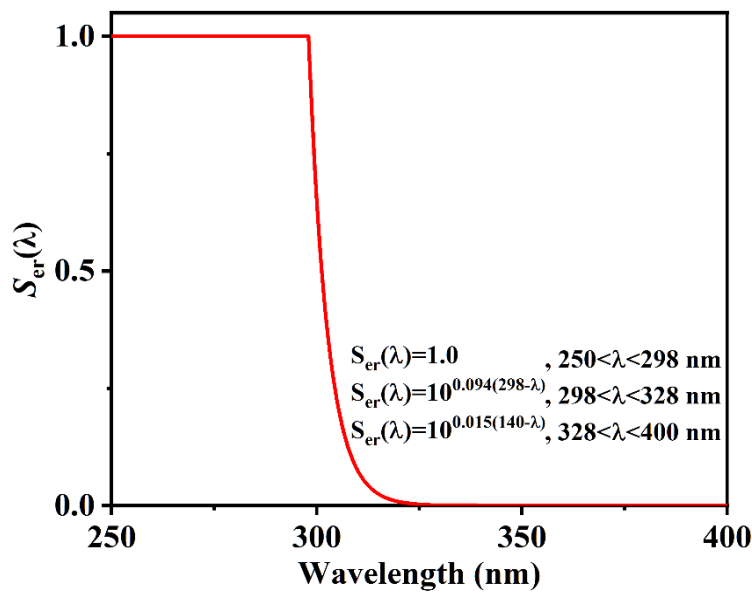

**Figure S8.** The plot of $S_{\text{er}}(\lambda)$.